\documentclass[floatfix,epsf,prb,twocolumn,amsmath,amssymb,showpacs,superscriptaddress]{revtex4}
\usepackage[T1]{fontenc}
\usepackage[english]{babel}
\usepackage{graphics}
\usepackage{amsmath}
\usepackage{amsfonts}
\usepackage{dsfont}
\usepackage{braket}
\usepackage{graphicx}
\usepackage{subfigure}
\usepackage{color}

\newcommand{\ph}{\varphi}

\renewcommand{\i}{\mathrm{i}}
\newcommand{\e}{\mathrm{e}}

\newcommand{\p}{\partial}
\DeclareMathAlphabet{\bi}{OML}{cmm}{b}{it}

\newcommand{\bsigma}{\boldsymbol{\sigma}}

\begin{document}

\title{Scattering of Dirac electrons by circular mass barriers:
valley filter and resonant scattering}
\author{M.~Ramezani Masir}\email{mrmphys@gmail.com}
\affiliation{Departement Fysica, Universiteit Antwerpen \\
Groenenborgerlaan 171, B-2020 Antwerpen, Belgium}
\author{A. Matulis}\email{amatulis@takas.lt}
\affiliation{Departement Fysica, Universiteit Antwerpen \\
Groenenborgerlaan 171, B-2020 Antwerpen, Belgium}
\affiliation{Semiconductor Physics Institute, Go\v{s}tauto 11,
LT-01108 Vilnius, Lithuania}
\author{F.~M.~Peeters}\email{francois.peeters@ua.ac.be}
\affiliation{Departement Fysica, Universiteit Antwerpen \\
Groenenborgerlaan 171, B-2020 Antwerpen, Belgium}
\date{Antwerpen, September 28, 2011}

\begin{abstract}
The scattering of two-dimensional (2D) massless Dirac electrons is investigated in the presence of a random array
of circular mass barriers. The inverse momentum relaxation time and the Hall factor
are calculated and used to obtain parallel and perpendicular resistivity components within linear transport
theory.  We found a non zero perpendicular resistivity component which has opposite
sign for electrons in the different $K$ and $K'$ valleys. This property can be used for
valley filter purposes. The total cross-section for scattering on penetrable barriers exhibit resonances
due to the presence of quasi-bound states in the barriers that show up as sharp gaps
in the cross-section while for Schr\"{o}dinger electrons they appear as peaks.
\end{abstract}
\maketitle

\section{Introduction}
Nanostructures have become the system of choice for studying transport over the past few
years\cite{Ferry}. Starting with 2D (two dimensional) electron systems at the interface
of two different materials several decades ago\cite{N0}, recently it has shifted to 2D
\emph{relativistic} materials e.~g.~graphene \cite{nov05,zang05,Nori1} and topological
insulators\cite{moo10, Hasan}. In pristine graphene the conduction and valence bands
touch each other in six points of the Brillouin zone and are defined by two independent sets
of cones commonly called $K$ and $K'$. Near these points the electronic dispersion is linear
which corresponds to the dispersion of massless relativistic particles, described by the
Dirac-Weyl  equation\cite{semenof,wall47}.

During the last decades there has been a lot of theoretical and experimental attempts to use
the spin of the electron as a carrier of information\cite{Wolf}. Graphene in addition to the spin of
the electron has two more degrees of freedom, sublattice pseudospin and valley isospin or valley index.
In order to scatter an electron from the $K$ valley to the $K'$ valley a large transfer of momentum is needed.
Typical disorder and Coulomb-type of scattering is unable to provide this momentum and in such a case
the valley isospin is a conserved quantum number in electronic transport. This allows us
to use valley isospin as a carrier of information. It was shown that graphene
nanoribbons with zigzag edges\cite{been1} can be used as a valley filter. Another promising
possibility to control the valley index of electrons is by using line defects\cite{mass3}. These can
be formed in graphene when grown on a Nickel substrate or by using so called mass barriers that
can be created by proper arrangement of dopants in the graphene sheet\cite{mass1,mass2}.

The purpose of this work is to apply the above mentioned ideas to electron transport
in the presence of circular mass barriers.
We solve the Dirac electron scattering problem on a sharp circular mass barrier and
calculated the cross-section, the inverse momentum relaxation time and the probability for the
electron to be reflected in the perpendicular direction. In spite of the circular symmetry of the
scatterers we obtain a non zero perpendicular component of resistivity that allows us to separate
electrons with different chirality, or belonging to $K$ and $K'$ valleys.
The scattering of Dirac electrons by a penetrable circular mass barrier
is influenced by the presence of quasi-bound states that results in resonant behavior.
The obtained results are compared with those for standard Schr\"{o}dinger electrons.

The paper is organized as follows. In section II we introduce the problem and illustrate our
formalism by considering first the more simple problem of Dirac electrons scattered by a circular hole.
In section III the boundary condition is obtained for a circular barrier i.~e.~a mass dot and a mass anti-dot.
The bound states in the mass dot are calculated in section IV. In section V  the scattering by an
impenetrable circle is considered and the total cross section, relaxation time and Hall component
of the resistivity is obtained. In section VI we repeat this calculation for a penetrable circle
and compared the results with the results for standard electrons that are presented in
Appendix A. Our conclusions are presented in section VII.

\section{Problem}

\begin{figure}[!ht]
  \begin{center}
  \includegraphics[width=8cm]{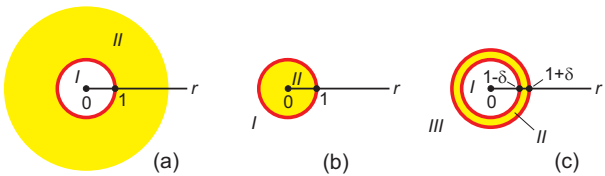}
  \end{center}
  \caption{(Color online) Layout of the mass barrier profile: (a) -- dot, (b) -- anti-dot, and
  (c) -- penetrable circle, i.~e~ a ring shaped barrier.}
  \label{fig1}
\end{figure}

We consider a Dirac electron interacting with circular barrier structures
shown in Fig.~\ref{fig1} by the shadowed regions.
In the long wave approximation it is described by the stationary equation
\begin{equation}\label{steq}
  \left\{H - E\right\}\Psi = 0
\end{equation}
with the following Dirac Hamiltonian:
\begin{equation}\label{dirtotham}
  H_D = \begin{pmatrix} \bsigma\cdot {\bf p} + \kappa\sigma_z & 0 \\
  0 & \bsigma^*\cdot {\bf p} - \kappa'\sigma_z
  \end{pmatrix}.
\end{equation}
This $4\times 4$ matrix Hamiltonian describes low energy excitations in $K$ and $K'$ valleys.
Due to its diagonal form it is possible to separate the scattering problems
in each $K$ or $K'$ valley. We choose the following $2\times 2$ Hamiltonians:
\begin{subequations}\label{valham}
\begin{eqnarray}
\label{valham1}
  H &=& \bsigma\cdot{\bf p} + \kappa({\bf r})\sigma_z, \\
\label{valham2}
  H' &=& \bsigma^*\cdot{\bf p} - \kappa'({\bf r})\sigma_z,
\end{eqnarray}
\end{subequations}
where $\bsigma=\{\sigma_x,\sigma_y\}$ and $\sigma_z$ stand for the Pauli matrices,
and $\kappa,({\bf r}),\,\kappa'({\bf r})$ characterizes the mass barrier for $K$
and $K'$ electrons, respectively.
We use dimensionless variables where velocities are measured in Fermi velocity
unit $v_F$, all coordinates are measured in the radius $R$ of the circular scattering barrier,
shown in Fig.~\ref{fig1} by the solid red line, and the electron energy --- in $\hbar v_F/R$ units.
From now on all equations are for $K$ valley electrons, except if otherwise specified.

According to standard scattering theory we present the wave function as
\begin{equation}\label{wfs}
  \Psi = \frac{1}{\sqrt{2}}\begin{pmatrix} 1 \\ 1 \end{pmatrix}e^{iEx}
  + \frac{f(\ph)}{\sqrt{r}}e^{iEr},
\end{equation}
consisting on the incident wave in the $x$-direction and the scattered part.
The scattered part is characterized by the two component scattering amplitude
\begin{equation}\label{fcomp}
  f(\ph) = \begin{pmatrix} A \\ B\end{pmatrix}.
\end{equation}
As the incoming wave function part is normalized to unit flow density
the differential cross-section is equal to the radial flow of electrons
corresponding to the scattered wave function part, namely,
\begin{equation}\label{dcur}
  \sigma(\ph) = f^+(\ph)\left(\bsigma\cdot{\bf n}\right)f(\ph)
  = \e^{i\ph}AB^* + \e^{-i\ph}A^*B,
\end{equation}
where ${\bf n}=(\cos\ph,\sin\ph)$ is the unit vector in the considered direction.

Besides the above differential cross-section we shall consider the
total cross-section
\begin{equation}\label{tot}
  \sigma = \int_0^{2\pi}d\ph\sigma(\ph),
\end{equation}
and two more averages: the inverse electron momentum relaxation time
(the quantity proportional to the dissipative component of resistivity)
\begin{equation}\label{irl}
  \gamma = \int_0^{2\pi}d\ph\sigma(\ph)\left(1 - \cos\ph\right),
\end{equation}
and the quantity
\begin{equation}\label{hr}
  \eta = \int_0^{2\pi}d\ph\sigma(\ph)\sin\ph,
\end{equation}
that corresponds to the perpendicular component of resistivity (or the
analog of the Hall component in the case with magnetic field).

Due to the azimuthal symmetry of our problems it is convenient to
use polar coordinates. Thus, assuming the total wave function as
\begin{equation}\label{wfsp}
  \Psi({\bf r}) = \begin{pmatrix} U({\bf r}) \\ V({\bf r}) \end{pmatrix},
\end{equation}
we shall solve the set of equations
\begin{subequations}\label{eqkc0}
\begin{eqnarray}
  e^{-\i\ph}\left(\frac{\p}{\p r} - \frac{\i}{r}\frac{\p}{\p\ph}\right)V &=& \i(E - \kappa)U, \\
  e^{\i\ph}\left(\frac{\p}{\p r} + \frac{\i}{r}\frac{\p}{\p\ph}\right)U &=& \i(E + \kappa)V
\end{eqnarray}
\end{subequations}
for the wave function components.

Now we expand the wave function into partial waves
\begin{equation}\label{wf}
  \begin{pmatrix} U({\bf r}) \\ V({\bf r}) \end{pmatrix}
  = \sum_{m=-\infty}^{\infty}w_m e^{im\ph}\begin{pmatrix} u_m(r) \\ ie^{i\ph} v_m(r) \end{pmatrix},
\end{equation}
replacing Eqs.~(\ref{eqkc0}) by the following set of two radial equations:
\begin{subequations}\label{eqkc1}
\begin{eqnarray}
  \left(\frac{d}{dr} + \frac{m+1}{r}\right)v_m &=& (E - \kappa)u_m, \\
  \left(\frac{d}{dr} - \frac{m}{r}\right)u_m &=& -(\kappa + E)v_m.
\end{eqnarray}
\end{subequations}
These equations are our main instrument when considering the problems shown in Fig.~\ref{fig1}.

\section{Boundary conditions}

For the sake of simplicity we restrict ourselves to model problems with large mass potentials,
and replace the potentials by proper boundary conditions on the electron wave functions. In fact, this
is the standard way of describing low energy scattering. For this purpose we have to
solve the appropriate equations in the barrier region, to apply the standard boundary conditions
for both wave function components on the solid circles shown in Fig.~\ref{fig1}, and to
calculate the limit $\kappa\to\infty$.
We start with the system shown in Fig.~\ref{fig1}(c).
Thus, we have to solve the following approximate set of radial equations:
\begin{subequations}\label{eqkc2}
\begin{eqnarray}
  \frac{d}{dr}v &=& - \kappa u, \\
  \frac{d}{dr}u &=& - \kappa v
\end{eqnarray}
\end{subequations}
in the thin shadowed region $II$ which is delimited by two circles of radius $1\pm\delta$
($\delta \ll 1$). Its solution reads
\begin{subequations}\label{dtsbc}
\begin{eqnarray}
  u &=& Fe^{\kappa(r-1)} + Ge^{-\kappa(r-1)}, \\
  v &=& -Fe^{\kappa(r-1)}  + Ge^{-\kappa(r-1)}.
\end{eqnarray}
\end{subequations}
Now satisfying the boundary conditions on the two circles demarcating the region $II$
we obtain the following set of four  algebraic equations
\begin{subequations}\label{ddbcbc}
\begin{eqnarray}
\label{ddbcbc1}
  u_I(1-\delta) &=& F\e^{-\kappa\delta} + G\e^{\kappa\delta}, \\
\label{ddbcbc2}
  v_I(1-\delta) &=& -F\e^{-\kappa\delta} + G\e^{\kappa\delta}, \\
\label{ddbcbc3}
  u_{III}(1+\delta) &=& F\e^{\kappa\delta} + G\e^{-\kappa\delta}, \\
\label{ddbcbc4}
  v_{III}(1+\delta) &=& -F\e^{\kappa\delta} + G\e^{-\kappa\delta}.
\end{eqnarray}
\end{subequations}
Eliminating the coefficients $F$ and $G$ we obtain in the limit $\delta\ll 1$
the boundary conditions

\begin{subequations}\label{ff}
\begin{eqnarray}
\label{ff1}
  v_{III}(1) - v_{I}(1) &=& -\left\{u_{III}(1) + u_{I}(1)\right\}\tanh(\kappa\delta),\phantom{mmu} \\
\label{ff2}
  u_{III}(1) - u_{I}(1) &=& -\left\{v_{III}(1) + v_{I}(1)\right\}\tanh(\kappa\delta),
\end{eqnarray}
\end{subequations}
connecting wave function components in the regions $I$ and $III$.

For a very thin and very high mass barrier (the analog of $\delta$-type barrier for
Schr\"{o}dinger electrons, see Sec.~\ref{sec_schrel}) we take the following limits:
\begin{equation}\label{dlim}
  \delta \to 0, \quad \kappa \to \infty, \quad \tanh(\kappa\delta) = P = \mathrm{const},
\end{equation}
what enables us to rewrite Eqs.~(\ref{ff}) as
\begin{subequations}\label{dbcff}
\begin{eqnarray}
\label{dbcff1}
  v_{III}(1) - v_{I}(1) &=& -P\left\{u_{III}(1) + u_{I}(1)\right\}, \\
\label{dbcff2}
  u_{III}(1) - u_{I}(1) &=& -P\left\{v_{III}(1) + v_{I}(1)\right\}.
\end{eqnarray}
\end{subequations}
These boundary conditions can be formally replaced by inserting Dirac $\delta$-functions
into Eqs.~(\ref{eqkc1}), namely, replacing those equations by
\begin{subequations}\label{eqkc3}
\begin{eqnarray}
  \left(\frac{d}{dr} + \frac{m+1}{r}\right)v_m &=& \left[E - 2P\delta(r-1)\right]u_m, \\
  \left(\frac{d}{dr} - \frac{m}{r}\right)u_m &=& -\left[2P\delta(r-1) + E\right]v_m,\phantom{mm}
\end{eqnarray}
\end{subequations}
if we assume the following rule for calculating the integrals when the integrand is a product of
Dirac $\delta$-function and some function $f(x)$ with the discontinuity\cite{roy93}:
\begin{equation}\label{dfr}
  \lim_{\alpha\to +0}\int_{-\alpha}^{\alpha}dx\delta(x)f(x) = \frac{1}{2}\left\{f(+0) + f(-0)\right\}.
\end{equation}
The parameter $P$ characterizes the effective strength of the $\delta$-type barrier,
and never exceeds the value $P=1$ in contrast to the case of Schr\"{o}dinger electrons where
it can take any value. A second major difference is that the wave function components
$u(r)$ and $v(r)$ are discontinuous at the position of the $\delta$-function (but the
probability density is continuous) while for Schr\"{o}dinger electrons the wave function
is continuous but the derivative of the wave function is discontinuous in that position.
This is a consequense of the fact that the Dirac-Weyl equation is a first order diferential
equation while the Schr\"{o}dinger equation is second order.

The obtained boundary conditions for the general case shown in Fig.~\ref{fig1}(c) enables us to
construct analogous boundary conditions for the two other cases
shown by Figs.~\ref{fig1}(a,b). So, in the case of Fig.~\ref{fig1}(a) we
assume
\begin{equation}\label{as1}
  \kappa = \infty, \quad P=1, \quad u_{III}(1) = v_{III}(1) = 0,
\end{equation}
and rewrite the boundary conditions given by Eqs.~(\ref{dbcff}) as
\begin{equation}\label{bcdot}
  u_{I}(1) = v_I(1).
\end{equation}
This is the boundary condition for the quantum dot, surrounded by
an infinite mass barrier.

In the case of Fig.~\ref{fig1}(b) an analogous reasoning leads to the following
boundary condition:
\begin{equation}\label{bcscatt}
  u_{I}(1) = -v_{I}(1),
\end{equation}
that we shall use for describing Dirac electron scattering by a hard wall anti-dot.

The obtained boundary conditions enables us to neglect the $\kappa$ terms in Eqs.~(\ref{eqkc1})
and solve the Dirac equations for free electrons
\begin{subequations}\label{eqkcf}
\begin{eqnarray}
  \left(\frac{d}{dr} + \frac{m+1}{r}\right)v_m &=& Eu_m, \\
  \left(\frac{d}{dr} - \frac{m}{r}\right)u_m &=& - Ev_m
\end{eqnarray}
\end{subequations}
in regions $I$ and $III$ separately.

\section{Bound states in the dot}
\label{sec_dot}

The most simple problem is the one of a quantum dot shown in Fig.~\ref{fig1}(a).
In this case the solution of Eqs.~(\ref{eqkcf}) in region $I$ has to be finite at
$r=0$ and is given by Bessel functions
\begin{subequations}\label{dsoldot}
\begin{eqnarray}
  u(r) &=& FJ_m(Er), \\
  v(r) &=& FJ_{m+1}(Er).
\end{eqnarray}
\end{subequations}
Satisfying boundary condition (\ref{bcdot}) we immediately arrive at
the algebraic equation
\begin{equation}\label{algequdot}
  J_m(E) - J_{m+1}(E) = 0
\end{equation}
that determines the energy of the bound states.

\section{Scattering by an impenetrable circle}
\label{sec_impen}

Now we shall consider our main problems, namely, the scattering of Dirac
electrons by circular mass barriers. We shall start with the case presented in Fig.~\ref{fig1}(b)
--- the scattering by an impenetrable circle (or by a hard wall anti-dot).
According to standard scattering theory
we have to construct the total wave function (\ref{wfsp}) solving the Dirac
equations for free electrons (\ref{eqkcf}) in the outer region $I$ that
satisfies the boundary condition (\ref{bcscatt}), to exclude the incoming part of the wave function,
and calculate the cross-section by means of Eq.~(\ref{dcur}).

So, the solution of Eqs.~(\ref{eqkcf}) in the outer region $I$ reads
\begin{subequations}\label{ksol}
\begin{eqnarray}
  u_m(r) = w_m\left[J_m(Er)\cos\delta_m + Y_m(Er)\sin\delta_m\right],\phantom{mmmm} \\
  v_m(r) = w_m\left[J_{m+1}(Er)\cos\delta_m + Y_{m+1}(Er)\sin\delta_m\right].\phantom{mm}
\end{eqnarray}
\end{subequations}
where the symbols $J_m$ and $Y_m$ stand for Bessel and Neumann functions, respectively.

Now satisfying the boundary conditions (\ref{bcscatt}) for any partial wave function harmonic
individually we obtain the expansion coefficients $\sin\delta_m$ and $\cos\delta_m$
through the so called phase shifts
\begin{equation}\label{scbcf}
  \tan\delta_m = -\frac{J_m(E) + J_{m+1}(E)}{Y_m(E) + Y_{m+1}(E)}.
\end{equation}

Usually the exclusion of the incoming plane wave from the total wave function (\ref{wf})
is done in the asymptotic region where $Er\gg 1$.
Here we use the asymptotic of the Bessel functions $J_m(Er) \approx \sqrt{2/\pi Er}\cos\Delta_m$,
$Y_m(Er) \approx \sqrt{2/\pi Er}\sin\Delta_m$ where
\begin{equation}\label{besselasymp3}
  \Delta_m = Er - \frac{\pi}{2}m - \frac{\pi}{4},
\end{equation}
which allows us to write the total wave function components as
\begin{subequations}\label{sdirass0}
\begin{eqnarray}
\label{sdirass1}
  U(\bi{r}) &=& \sqrt{\frac{2}{\pi kr}}\sum_{m=-\infty}^{\infty}w_m e^{i m\ph}
  \cos(\Delta_m-\delta_m), \\
\label{sdirass2}
  V(\bi{r}) &=& i\sqrt{\frac{2}{\pi kr}}\sum_{m=-\infty}^{\infty}w_m e^{i(m+1)\ph}
  \cos(\Delta_{m+1}-\delta_m).\phantom{mmm}
\end{eqnarray}
\end{subequations}
In this asymptotic region the incoming plane wave can be presented as
\begin{equation}\label{asymptcyl}
\begin{split}
  \e^{iEx} &= \sum_{m=-\infty}^{\infty} i^m e^{im\ph}J_m(Er) \\
  &\approx \sqrt{\frac{2}{\pi Er}}\sum_{m=-\infty}^{\infty}i^me^{im\ph}\cos\Delta_m \\
  &= \sqrt{\frac{1}{2\pi Er}}\sum_{m=-\infty}^{\infty}i^me^{im\ph} \left(
  e^{i\Delta_m} + e^{-i\Delta_m}\right).
\end{split}
\end{equation}
Now in order to compensate the incoming term in the total wave function (\ref{wf}) by the
second term of the last line in Eq.~(\ref{asymptcyl}) we have to take
\begin{equation}\label{phase}
  w_m = \frac{1}{\sqrt{2}}i^me^{-i\delta_m}.
\end{equation}
It is remarkable that this choice makes the above mentioned compensation
in both scattering amplitude components $A$ and $B$, but not in every pair
of radial components $u_m$ and $v_m$ separately.
This is related to the fact that the angular momentum is not conserved quantity but that
the Dirac-Weyl Hamiltonian commutes with operator consisting of the angular momentum
plus isospin. So, inserting the above expression of
$w_m$ into Eq.~(\ref{sdirass0}) and subtracting the incoming wave function part
given by Eq.~(\ref{asymptcyl}) we arrive after laborious but straightforward calculations
at the following expression for the components of the scattering amplitude (\ref{fcomp}):
\begin{subequations}\label{pwsc}
\begin{eqnarray}
  A &=& e^{-3i\pi/4}\sqrt{\frac{1}{\pi E}}\sum_{m=-\infty}^{\infty}
  e^{im\ph}e^{-i\delta_m}\sin\delta_m, \\
  B &=& e^{-3i\pi/4}e^{i\ph}\sqrt{\frac{1}{\pi E}}\sum_{m=-\infty}^{\infty}
  e^{im\ph}e^{-i\delta_m}\sin\delta_m.\phantom{mm}
\end{eqnarray}
\end{subequations}
Inserting them into Eq.~(\ref{dcur}) we obtain the following differential cross-section:
\begin{equation}\label{diffcrossfin}
  \sigma(\ph) = \frac{2}{\pi E}\sum_{m,m'=-\infty}^{\infty}e^{i\left[(m-m')\ph-(\delta_m-\delta_{m'})\right]}
  \sin\delta_m\sin\delta_{m'}.
\end{equation}

We have to keep in mind that the derivation of this cross-section was performed
with the Hamiltonian (\ref{valham1}), which is valid for electrons
in the $K$ valley.

To obtain the results for $K'$ valley electrons we have
to repeat the procedure starting with Hamiltonian (\ref{valham2}). Fortunately,
it leads to the same set of differential equations (\ref{eqkc0}) with a single
change $U\rightleftarrows V$. Further, we replace Eq.~(\ref{wf}) by
\begin{equation}\label{pwf}
  \begin{pmatrix} U({\bf r}) \\ V({\bf r}) \end{pmatrix}
  = \sum_{m=-\infty}^{\infty}w_m e^{-im\ph}\begin{pmatrix} u_m(r) \\ -ie^{-i\ph} v_m(r) \end{pmatrix},
\end{equation}
and arrive at the same radial equation set (\ref{eqkc1}) as was obtained for
the $K$ valley case. Consequently, all equations, including the boundary condition, remain
the same for the $K'$ valley as well. The equation for the differential cross-section
for the $K'$ valley, however, differs. In this case the changes $m\to-m$ and $m'\to-m'$
have to be performed in the argument of the exponent leaving the same indexes of
phase shifts $\delta_m$ and $\delta_{m'}$.

Now inserting the obtained differential cross-section (\ref{diffcrossfin}) into Eq.~(\ref{tot})
we obtain the total cross-section
\begin{equation}\label{totsig}
\begin{split}
  \sigma &= \frac{4}{E}\sum_{m,m'=-\infty}^{\infty}\delta_{m,m'}e^{i(\delta_{m'}-\delta_m)}
  \sin\delta_m\sin\delta_{m'} \\
  &= \frac{4}{E}\sum_{m=-\infty}^{\infty}\sin^2\delta_m.
\end{split}
\end{equation}
Note the above mentioned change $m\to-m$ and $m'\to -m'$ in the exponent argument now
appears as the same change in the argument of the Kronecker symbol $\delta_{m,m'}$.
It is evident that this change does not influence the total cross-section, and
consequently, it is the same for both $K$ and $K'$ valley electrons.

Taking into account Eq.~(\ref{scbcf}) the partial cross-section contribution to the
total cross-section can be presented as
\begin{equation}\label{totpar}
\begin{split}
  &  \sin^2\delta_m \\
  &= \frac{\left[J_m(E) + J_{m+1}(E)\right]^2}
  {\left[J_m(E) + J_{m+1}(E)\right]^2 + \left[Y_m(E) + Y_{m+1}(E)\right]^2},
\end{split}
\end{equation}
what enables us to calculate the scattering cross-section directly.

\begin{figure}[!ht]
  \begin{center}
  \includegraphics[width=8cm]{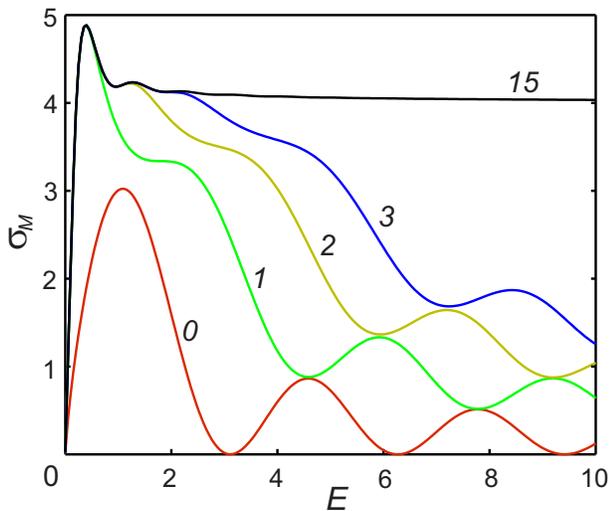}
  \end{center}
  \caption{(Color online) The energy dependence of the partial sums $\sigma_M$ contributing
  to the total cross-section. The italic numbers on the curves indicate the number $M$.
  The black curve for $M=15$ corresponds to the convergent result.}
  \label{fig2}
\end{figure}

The energy dependence of the total cross-section where the sum is restricted by the value $M$
($|m|\leqslant M$) is shown in Fig.~\ref{fig2}.
We see a rather good convergence at low energies where three terms (i.~e.~$m=0,\pm 1$)
are already sufficient.

The oscillating behavior of the energy dependence of the partial sum follows from the same
behavior of the separate terms in Eq.~(\ref{totsig}) that can be easily explained calculating
the asymptotic of the phase shifts. Indeed, using the asymptotic of the Bessel functions
and replacing $\Delta_m$ by $\Delta_m|_{r=1}$ we
have the following asymptotic expression for the scattering phase (\ref{scbcf}):
\begin{equation}\label{di}
\begin{split}
  \tan\delta_m &= -\frac{J_m(E) + J_{m+1}(E)}{Y_m(E) + Y_{m+1}(E)}
  \approx \frac{\cos\Delta_m + \cos\Delta_{m+1}}{\sin\Delta_m + \sin\Delta_{m+1}} \\
& = \frac{\cos\Delta_m + \sin\Delta_m}{\sin\Delta_m - \cos\Delta_m}
  = -\frac{\sin\left(\Delta_m + \pi/4\right)}{\cos\left(\Delta_m + \pi/4\right)} \\
&  = -\tan\left(\Delta_m + \pi/4\right).
\end{split}
\end{equation}
Thus, in the asymptotic region we have
\begin{equation}\label{phas}
  \sin^2\delta_m = \sin^2(E - \pi m/2),
\end{equation}
what explains the waving behavior of the obtained partial contributions to the total cross-section.

By the way, this simple expression for the scattering phase enables us to perform the approximate summation
in Eq.~(\ref{totsig}) for large energies which results in the limit cross-section $\sigma_{\mathrm{lim}} = 4$
as can also be seen clearly in Fig.~\ref{fig2}.
This value is twice larger than the classical value $\sigma_{\mathrm{cl}}=2$ that can be
obtained assuming that relativistic electrons are moving along trajectories given by
non quantum mechanical equations of motion. This discrepancy is caused by the diffraction
of the electronic waves when they are scattered by hard wall type potentials, and it is inherent
over scattering by small angles. It is remarkable that relativistic electrons exhibit the same
feature as Schr\"{o}dinger electrons (see the textbook \cite{sakur94}).

In Appendix A we present similar results for Schr\"{o}dinger
electrons (see Fig.~\ref{fig9}). Note that there are some differences in their $k$-dependence:
1) in the low energy limit the cross-section of the Schr\"{o}dinger electrons diverge logarithmically,
while for Dirac electrons it becomes zero, and 2) $\sigma(k)$ for Schr\"{o}dinger electrons
is an uniform decreasing function of $k$ while for Dirac electrons it exhibits oscillations in the
low energy region. Both cross-sections approach the high energy limit from above.

Inserting differential cross-section (\ref{diffcrossfin}) into Eq.~(\ref{irl}) we
obtain the inverse momentum relaxation time (or the dissipative resistivity component)
\begin{equation}\label{pr}
\begin{split}
  \gamma =& \frac{2}{E}\sum_{m,m'=-\infty}^{\infty}
  \left[2 - \delta_{m',m+1} - \delta_{m',m-1}\right] \\
  &\phantom{mmmmm}\times e^{i(\delta_{m'}-\delta_m)}
  \sin\delta_m\sin\delta_{m'} \\
  =& \frac{2}{E}\sum_{m=-M}^{M}\sin^2\left(\delta_m-\delta_{m+1}\right).
\end{split}
\end{equation}
By the way the change $m\to-m$ and $m'\to-m'$ in the arguments of the Kronecker symbol
does not influence the value of the above expression. Consequently, the above inverse momentum
relaxation time expression is the same for both $K$ and $K'$ valley electrons.

\begin{figure}[!ht]
  \begin{center}
  \includegraphics[width=8cm]{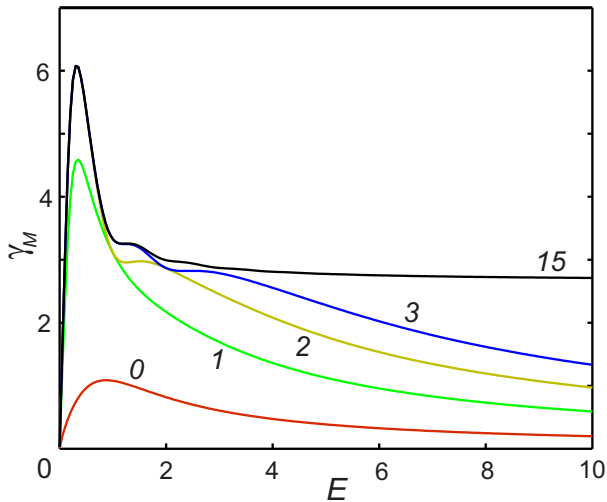}
  \end{center}
  \caption{(Color online) The contribution of the partial sums $\gamma_M$
  to the inverse momentum relaxation time. The italic numbers on the
  curves indicate the number $M$ where $M=15$ corresponds to the convergent result.}
  \label{fig3}
\end{figure}

Now inserting the phases obtained by solving Eq.~(\ref{scbcf}) into Eq.~(\ref{pr}) we get the result that
is shown in Fig.~\ref{fig3}.
It's behavior is qualitatively similar to the one for the total cross-section.
For large energies it approaches the limiting
value $\gamma=8/3$, which we obtain by calculating the integral (\ref{irl}) with the
classical differential cross-section, confirming the known fact that the integral (\ref{irl})
is not sensitive to forward scattering. Thus this relaxation time isn't affected by the above
mentioned discrepancy between the quantum and classical result as it was with the total cross-section.

Note that there is an essential difference in the separate contributions to the total cross-section
$\sigma$ and the inverse momentum relaxation time $\gamma$. The partial contributions to $\gamma$
do not exhibit any oscillating behavior that was inherent in the case of $\sigma$.
This is expected from Eq.~(\ref{di}) where the difference of neighboring phases
$(\delta_m-\delta_{m+1})$ doesn't depend on energy in the asymptotic region.

\begin{figure}[!ht]
  \begin{center}
  \includegraphics[width=8cm]{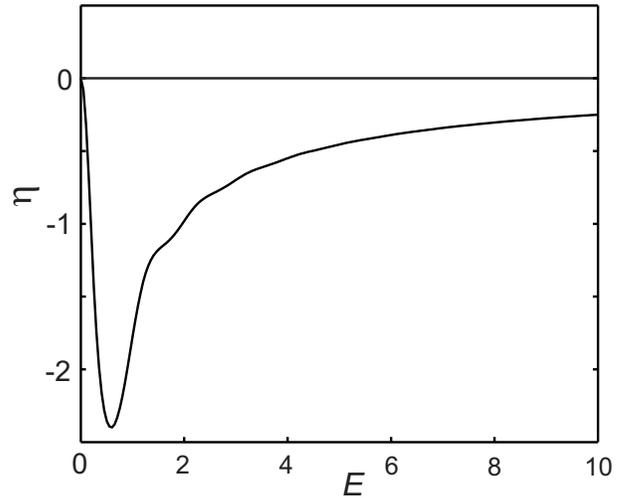}
  \end{center}
  \caption{The Hall component of resistivity.}
  \label{fig4}
\end{figure}

And at last inserting the differential cross-section (\ref{diffcrossfin}) into Eq.~(\ref{hr})
we obtain the perpendicular (or Hall) component of the conductivity
\begin{equation}\label{hrf}
\begin{split}
  \eta =& -\frac{2i}{\pi E}\sum_{m,m'=-\infty}^{\infty}
  \left[\delta_{m',m+1} - \delta_{m',m-1}\right] \\
  &\phantom{mmmmmm}\times e^{i(\delta_{m'}-\delta_m)}
  \sin\delta_m\sin\delta_{m'} \\
  =& \frac{4}{E}\sum_{m=-\infty}^{\infty}\sin\delta_m\sin\delta_{m+1}
  \sin\left(\delta_{m+1}-\delta_m\right).
\end{split}
\end{equation}
The result is shown in Fig.~\ref{fig4}.
Naively we would expect that $\eta=0$ at zero magnetic field.
To our surprise we find that $\eta<0$ and that it conserves it's
sign as a function of $E$. It means that the mass barrier acts similar as a magnetic field.
In order to obtain the result for $K'$ valley electrons we have to change $m\to-m$ and $m'\to-m'$
in the arguments of Kronecker symbols in the first line of Eq.~(\ref{hrf}).
It is evident that due to this change the Hall component of the resistivity $\eta$ changes it's sign.
Thus, the electrons from different valleys are deflected to opposite sides of the sample.
There is no net charge build up across the sample and thus no Hall voltage. But there is
a separation of different $K$ and $K'$ valley electrons across the sample and thus we
can use this effect for valley filtering purposes.

\section{Scattering on a penetrable circle}

Now we turn to our last problem --- scattering of Dirac electrons by a penetrable circle
in order to demonstrate how possible quasi-bound states reveal themselves in the
scattering cross-section. For this purpose we have to solve the Dirac equation for
free electrons (\ref{eqkcf}) in both $I$ and $III$ regions and to apply boundary conditions
(\ref{dbcff}). These solutions are given by Eq.~(\ref{dsoldot}) for
the inner region $I$, and by Eq.~(\ref{ksol}) for the outer region $III$.
Moreover, the procedure of the exclusion of the incident exponent is the same as it
was performed in Sec.~\ref{sec_impen}. Thus we can immediately write down
Eqs.~(\ref{pwsc}) for the components of the scattering amplitudes, and use the previous expressions
for the scattering cross-section (\ref{diffcrossfin},\ref{totsig},\ref{pr},\ref{hrf}).

The single procedure that should be performed is to satisfy the boundary conditions
(\ref{dbcff}) and calculate the phase shifts $\delta_m$. Inserting into Eq.~(\ref{dbcff}) the
solutions (\ref{dsoldot}) and (\ref{ksol}) we obtain the set of two equations:
\begin{subequations}\label{dscpen}
\begin{eqnarray}
\label{dbcf1}
  && CJ_{m+1} + SY_{m+1} - FJ_{m+1} \nonumber \\
  && = -P\left\{CJ_m + SY_m + FJ_m\right\}, \\
\label{dbcf2}
  && CJ_m + SY_m - FJ_m \nonumber \\
  && = -P\left\{CJ_{m+1} + SY_{m+1} + FJ_{m+1}\right\},
\end{eqnarray}
\end{subequations}
where for the sake of shortness we denoted
\begin{equation}\label{dencs}
  C = w_m\cos\delta_m, \quad S = w_m\sin\delta_m,
\end{equation}
and omitted the arguments $E$ of all Bessel functions.

Now excluding coefficient $F$ we obtain a single equation.
It can be solved for the tangent of the phase shift, and
using the expression for the Wronskian of the Bessel functions we arrive at
\begin{equation}\label{dsf3}
  \tan\delta_m = -\frac{(J_{m+1} + J_m)(J_m - J_{m+1})}
  {(Y_{m+1} + Y_m)(J_m - J_{m+1}) - 1/pE},
\end{equation}
where the symbol
\begin{equation}\label{symbp}
  p = \frac{\pi P}{(1-P)^2}
\end{equation}
characterizes the impenetrability of the circle. The value $p=\infty$ corresponds to a completely
impenetrable circle, i.~e.~the previously considered scattering on an impenetrable
anti-dot, while the value $p=0$ corresponds to the case of complete penetration, or
the absence of any scatterer.

\begin{figure}[!ht]
  \begin{center}
  \includegraphics[width=8cm]{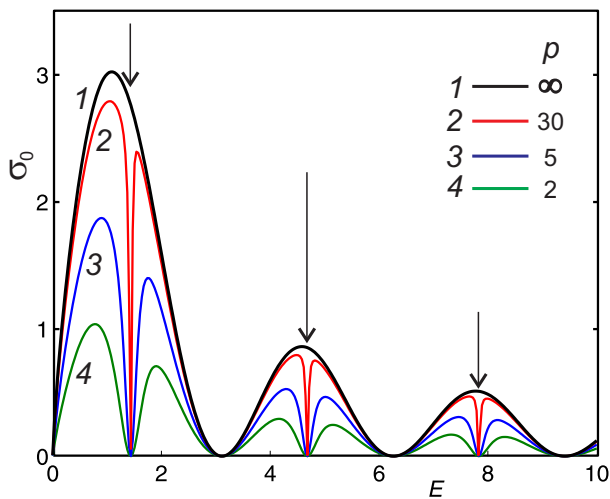}
  \end{center}
  \caption{(Color online) The partial $\sigma_0$ contribution to the total cross-section
  for different values of the penetrable parameter $p$ of the circular scatterer.
  The curve marked by $p=\infty$ corresponds to the case of scattering by an impenetrable
  circle.}
  \label{fig5}
\end{figure}

The numerical results for the lowest contribution ($m=0$) to the total cross-section
are shown in Fig.~\ref{fig5} for different $p$ values.
The vertical lines indicate the energies of the bound states of the dot obtained by
solving Eq.~(\ref{algequdot}) as described in Sec.~\ref{sec_dot}.
Although now the dot is penetrable and it has no bound states, the corresponding
quasi-bound states reveal themselves as narrow gaps close to the maxima of the oscillating
partial contribution. Note that this is a particular feature of Dirac electrons. While in the
case of Schr\"{o}dinger electrons the quasi-bound states appear as peaks in the
cross-section (see Figs.~\ref{fig10} and \ref{fig11} in Appendix A).

According to Eq.~(\ref{dsf3}) it seems that there should be one more set of gaps
in the partial cross-section, related to the equation
\begin{equation}\label{morepe}
  J_m(E) + J_{m+1}(E) = 0.
\end{equation}
But in this case, however, after neglecting the last term in the denominator of Eq.~(\ref{dsf3})
it coincides with the phase (\ref{scbcf}) obtained for the scattering by the
impenetrable circle, and in such a way it indicates that the above condition defines
the flat minimum related to the diffraction pattern in the cross-section considered
in Sec.~\ref{sec_impen}.

\begin{figure}[!ht]
  \begin{center}
  \includegraphics[width=8cm]{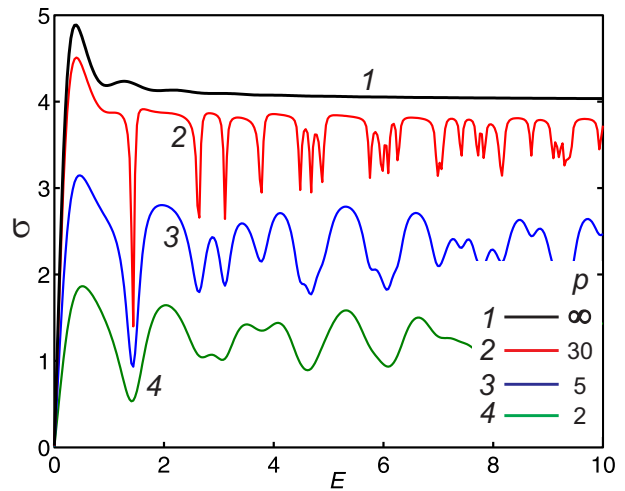}
  \end{center}
  \caption{(Color online) The total cross-section $\sigma$ for a penetrable circular
  scatterer for different $p$ values.
  The curve marked by $p=\infty$ corresponds to the case of scattering on an impenetrable
  circle.}
  \label{fig6}
\end{figure}

These gaps in the partial cross-section reveal themselves in the total cross-section as
shown in Fig.~\ref{fig6}.
Because of the contribution of the other partial waves the gaps no longer reach zero
as e.~g.~shown in the case of the partial cross-section $\sigma_0$.
Notice that they become more pronounced and narrower
when the parameter $p$ increases (when the circle becomes less penetrable).

\begin{figure}[!ht]
  \begin{center}
  \includegraphics[width=8cm]{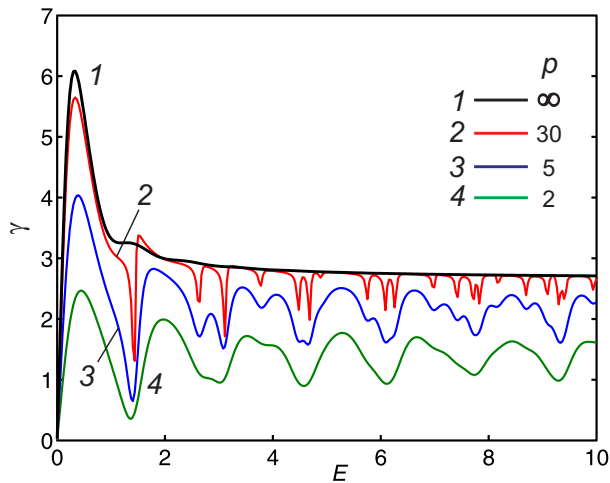}
  \end{center}
  \caption{(Color online) The inverse momentum relaxation time calculated according Eq.~(\ref{pr})
  for a penetrable circular scatterer.}
  \label{fig7}
\end{figure}

\begin{figure}[!ht]
  \begin{center}
  \includegraphics[width=8cm]{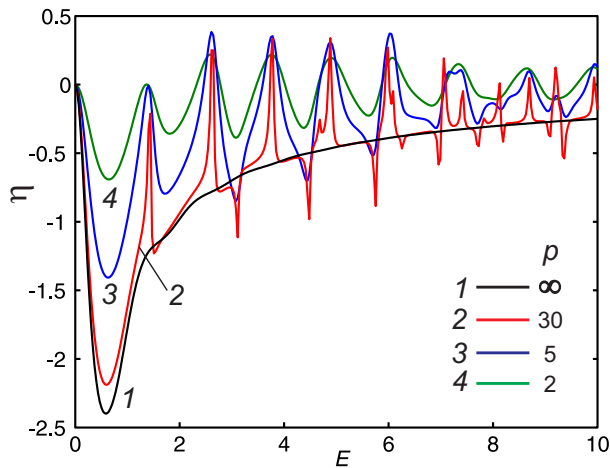}
  \end{center}
  \caption{(Color online) The Hall component $\eta$ calculated according Eq.~(\ref{hrf})
  for a penetrable circular scatterer.}
  \label{fig8}
\end{figure}

In Fig.~\ref{fig7} the results for the inverse momentum relaxation time
and in Fig.~\ref{fig8} those for the perpendicular component of the resistivity are presented.
We see that although Eqs.~(\ref{pr}) and (\ref{hrf}) are more sophisticated functions of the phases
$\delta_m$ and even two neighboring phases $\delta_m$ and $\delta_{m+1}$ are intermixed,
nevertheless the resonant behavior, i.~e.~the negative peaks, is still clearly visible
in the inverse momentum relaxation time. The behavior of peaks in the Hall component
of the resistivity, however, is more complicated, and exhibits both sharp peaks and dips
where now the sign of $\eta$ can change in small regions of energy.

\section{Conclusions}

We investigated the scattering of Dirac electrons by sharp circular mass barriers, where
we studied both hard wall type anti-dot and circular penetrable scatterer.
For this purpose the proper boundary conditions for Dirac equation were derived,
and it was illustrated how it is possible to use formally $\delta$-type
functions describing relativistic systems with high and sharp potentials.

The differential and total cross-section, the inverse momentum relaxation time
and the perpendicular (Hall) component of resistivity were calculated.
The obtained results were compared with analogous results for scattering
of Schr\"{o}dinger electrons by similar scatterers.

It was shown that the scattering of
Dirac electrons even by azimuthal symmetric structures depends on
the valley index: the  $K$ valley electrons are preferentially deflected
to one side (the Hall component of the resistivity isn't zero) while the electrons
of the other $K'$ valley are deflected to the other side of the sample. This enables
one to use this property for valley index filtering in transport experiments.

There is an essential difference in the energy dependence of the cross-section between
Dirac and Schr\"{o}dinger electrons. At small energies
the cross-section for Dirac electrons tend to zero while those for Schr\"{o}dinger
electrons diverge logarithmically.
This feature of Dirac electron is caused by the fact
that zero energy for Dirac electron actually corresponds to the middle of the half-filled band,
and not to the bottom of it as in the Schr\"{o}dinger electron case.

We showed that in the case of Dirac electron scattering on a penetrable circle the
quasi-bound states reveal themselves as sharp gaps in the total cross-section,
the inverse momentum relaxation time and the Hall component of the resistivity as well.
In the case of Schr\"{o}dinger electrons those resonances show up as peaks in the cross-sections,
and thus their appearance is qualitatively very different.

\begin{acknowledgments}
This work was supported by the European Science Foundation (ESF) under the EUROCORES
Program EuroGRAPHENE within the project CONGRAN.
\end{acknowledgments}

\appendix

\section{Scattering of a Schr\"{o}dinger electron by circular barriers}
\label{sec_schrel}

In this Appendix we study the scattering of Schr\"{o}dinger
electrons by sharp circular potentials shown in
Fig.~\ref{fig1}. This allows us to compare them with results for Dirac
electrons considered in the present paper.

Here scattering is now described by the single component wave function
$\Psi({\bf r})$ satisfying the stationary Schr\"{o}dinger equation
\begin{equation}\label{sc11}
    (\nabla^{2} + k^{2})\Psi(\bi{r}) =0,
\end{equation}
where the symbol $k$ stands for the electron momentum related to it's energy as
$E = k^2/2$. Now we use dimensionless units defined as follows. The coordinates
will be measured in the radius of the circular scatterer $R$, energy --- in $\hbar^2/mR^2$
units, and the electron momentum --- in $\hbar/R$ units.

In polar coordinates the wave function $\Psi({\bf r})$ is usually expanded into a Fourier series
like $U({\bf r})$ component of Dirac function in Eq.~(\ref{wf}) with radial components $\psi_m(r)$
satisfying the Bessel equation. That is why all mathematics is practically the same as
used in previous sections with a single replacement of the Dirac electron energy $E$ by the
momentum $k$ of the Schr\"{o}dinger electron.
Consequently, we obtain the same differential cross-section given by
 Eq.~(\ref{diffcrossfin}), and Eqs.~(\ref{totsig}) and (\ref{pr})
for the total cross-section and inverse momentum relaxation time.

Nevertheless, there is an essential difference between Dirac and Schr\"{o}dinger electrons which
is the different boundary conditions that will lead to different scattering phases.

\begin{figure}[!ht]
  \begin{center}
  \includegraphics[width=8cm]{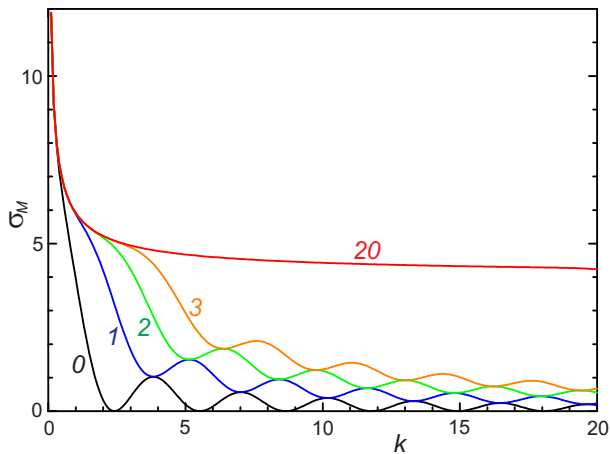}
  \end{center}
  \caption{(Color online) The energy dependence of the partial sums $\sigma_M$ contributing
  to the total cross-section for Schr\"{o}dinger electrons.
  The italic numbers on the curves indicate the number $M$.}
  \label{fig9}
\end{figure}

Thus, in the case of scattering on a circular hard wall potential (see Fig.~\ref{fig1}(a))
every radial component of the wave function has to satisfy the zero boundary condition
\begin{equation}\label{srzbc}
  \psi_m(1) = 0,
\end{equation}
what leads to the following equation for the phase
\begin{equation}\label{srphequ}
  \tan\delta_m = -\frac{J_{m}(ka)}{Y_{m}(ka)},
\end{equation}
instead of Eq.~(\ref{scbcf}) for the Dirac electron. The partial contributions and
the total cross-section calculated by using the above phase equation are shown in Fig.~\ref{fig9}.

\begin{figure}
  \begin{center}
  \includegraphics[width=8cm]{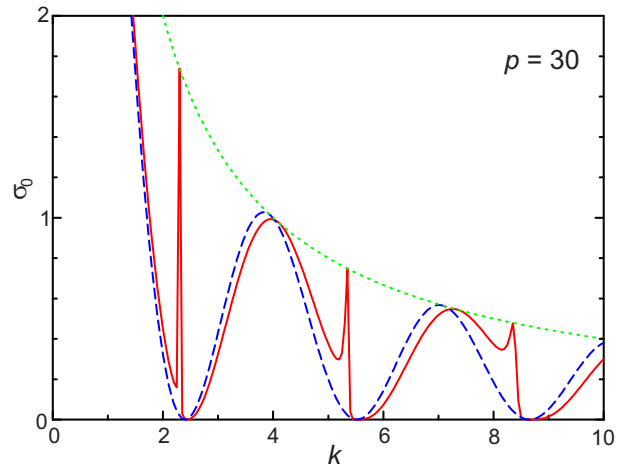}
  \end{center}
  \caption{(Color online) The $m=0$ contribution  to the total cross-section for Schr\"{o}dinger
  electrons scattered on penetrable circular potentials, shown by red solid curve. Green dotted curve is
  the envelop function $4/k$, and the blue dashed curve is the result for scattering on impenetrable scatterers.}
  \label{fig10}
\end{figure}

In the case of an extremely narrow penetrable circle, i.~e.~a $\delta$-type potential,
Eq.~(\ref{sc11}) has to be replaced by
\begin{equation}\label{plg}
    \left\{\nabla^{2} + k^{2} - p\delta(r-1)\right\}\Psi(\bi{r}) =0
\end{equation}
what leads to the following boundary conditions for the radial wave function components
on the circle:
\begin{subequations}\label{pks}
\begin{eqnarray}
  \psi_m(1+0) &=& \psi_m(1-0), \\
  \psi'_m(1+0) - \psi'_m(1-0) &=& p\psi_m(1).
\end{eqnarray}
\end{subequations}
This leads to the following scattering phase equation:
\begin{equation}\label{pfaz}
  \tan\delta_m = -\frac{J_m(k)}{Y_m(k) - 2/p\pi J_m(k)}.
\end{equation}
From it we obtain the following partial contribution to the total cross-section:
\begin{equation}\label{pparc}
  \frac{4}{k}\sin^2\delta_m = \frac{4J_m^2(k)}{k\left\{J_m^2(k) + [Y_m(k) - 2/p\pi J_m(k)]^2\right\}}.
\end{equation}
The typical contribution(when $m=0$) is shown in Fig.~\ref{fig10} by red solid curve.
Narrow peaks appear close to the positions of the bound states of a dot
that are defined by the equation $J_m(k)=0$.
This can be also seen from Eq.~(\ref{pparc}) which formally is similar to a Lorentzian curve.
The top of the peak is achieved when the second term in the denominator
(the analog of detuning) in Eq.~(\ref{pparc}) is zero. In the case of small
penetrability of the scatterer this can be realized if the large parameter $p\gg 1$
is compensated by a small $J_m(k) \ll 1$ value. But then the contribution becomes
equal to $4/k$ what indicates that the maximum of all peaks reach the above envelope
function shown by green dotted curve in Fig.~\ref{fig10}. One more property of
the partial contribution follows from the fact that it is rather close to
the same contribution in the case of scattering by impenetrable scatterer
shown by the blue dashed curve (Eq.~(\ref{pfaz}) converts itself into Eq.~(\ref{srphequ})
when $p=\infty$) what indicates that the peaks appear at the minima of that dashed
curve.

\begin{figure}
  \begin{center}
  \includegraphics[width=8cm]{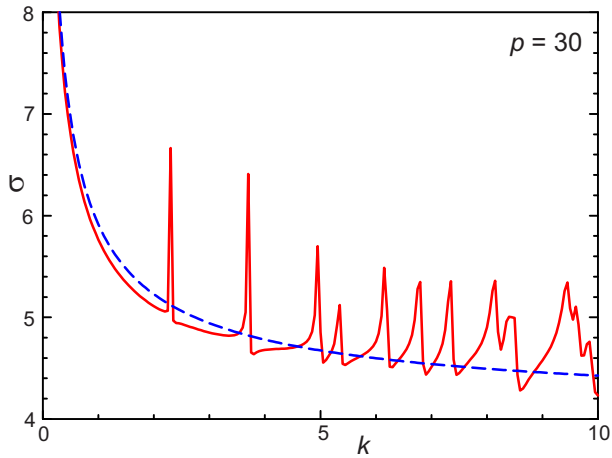}
  \end{center}
  \caption{(Color online) The same as Fig.~\ref{fig10} but now for the total cross-section.}
  \label{fig11}
\end{figure}

These resonances show up also in the total cross-section as seen in Fig.~\ref{fig11}.
The comparison with the blue dashed curve calculated for $p=\infty$ indicates clearly
that the quasi-bound states appear in the total cross-section as positive peaks.

\end{document}